\documentclass[conference]{IEEEtran}
\IEEEoverridecommandlockouts
\usepackage{cite}
\usepackage{float}
\usepackage{amsmath,amssymb,amsfonts}
\usepackage{algorithmic}
\usepackage{graphicx}
\usepackage{textcomp}
\usepackage{xcolor}

\def\BibTeX{{\rm B\kern-.05em{\sc i\kern-.025em b}\kern-.08em
    T\kern-.1667em\lower.7ex\hbox{E}\kern-.125emX}}
\begin{document}

\title{Nonlinear Regression with a Convolutional \\ Encoder-Decoder for Remote Monitoring of Surface Electrocardiograms}

\author{
\IEEEauthorblockN{Anton Banta$^{1}$, Romain Cosentino$^{1}$, Mathews M John$^2$, Allison Post$^2$}
\IEEEauthorblockN{Skylar Buchan$^2$, Mehdi Razavi$^3$, Behnaam Aazhang$^1$}
\IEEEauthorblockA{$^1$\textit{Department of Electrical and Computer Engineering, Rice University}}
\IEEEauthorblockA{$^2$\textit{Electrophysiology Clinical Research and Innovations, Texas Heart Institute}}
\IEEEauthorblockA{$^3$\textit{Department of Cardiology, Texas Heart Institute}}
}

\maketitle

\begin{abstract}
We propose the Nonlinear Regression Convolutional Encoder-Decoder (NRCED), a novel framework for mapping a multivariate input to a multivariate output. In particular, we implement our algorithm within the scope of 12-lead surface electrocardiogram (ECG) reconstruction from intracardiac electrograms (EGM) and vice versa. The goal of performing this task is to allow for improved point-of-care monitoring of patients with an implanted device to treat cardiac pathologies. We will achieve this goal with 12-lead ECG reconstruction and by providing a new diagnostic tool for classifying atypical heartbeats. The algorithm is evaluated on a dataset retroactively collected from 14 patients. Correlation coefficients calculated between the reconstructed and the actual ECG show that the proposed NRCED method represents an efficient, accurate, and superior way to synthesize a 12-lead ECG. We can also achieve the same reconstruction accuracy with only one EGM lead as input. We also tested the model in a non-patient specific way and saw a reasonable correlation coefficient. The model was also executed in the reverse direction to produce EGM signals from a 12-lead ECG and found that the correlation was comparable to the forward direction. Lastly, we analyzed the features learned in the model and determined that the model learns an overcomplete basis of our 12-lead ECG space. We then use this basis of features to create a new diagnostic tool for identifying atypical and diseased heartbeats. This resulted in a ROC curve with an associated area under the curve value of 0.98, demonstrating excellent discrimination between the two classes.
\end{abstract}


\section{Introduction}
Heart failure affects over 5.8 million people in the USA and over 23 million worldwide \cite{bui2011epidemiology}. It is characterized as a dysfunction in the heart that leads to decreased blood flow to the body. A subset of these patients will require a pacemaker to treat the cardiac pathology. According to \cite{wood2002cardiac} there are about 3 million people worldwide with pacemakers, and each year 600,000 pacemakers are implanted. 

These patients require regular in-hospital visits to follow-up their response to the therapy. This entails checking if the pacemaker is working properly and perhaps modifying the pacing parameters. These parameters often need to be altered over time because the heart changes its physiology in response to the implantable device. More recently, wireless remote monitoring of the pacemaker has become a priority for all major pacemaker manufacturers. This allows the cardiologist to follow up on the efficacy of the treatment often without needing more hospital visits, which can be quite costly and time consuming. 

The pacemaker is sensing the heart's electrical conductance locally through electrodes placed on the endocardium and/or the epicardium. These signals are called intracardiac electrograms (EGM). While EGMs provide a vast amount of information, 12-lead electrocardiograms provide a wealth of information that allows for better clinical intervention. The full 12-lead ECG can only be obtained at the clinic using skin electrodes. The synthesis, or reconstruction, of the surface ECG from a set of EGM is thus of critical importance in this context. The problem of reconstructing the ECG from a set of EGM lies within the scope of multivariate multiple regression. That is, the input and the output are multivariate time-series. In the present case, the input is usually the EGM, a multi channel time-series (usually 2-5 leads) and the output the ECG, a $12$-channel time-series. In this work, we also consider the case where we regress the $12$-lead ECG to obtain the EGMs. In fact, this would provide a new tool for cardiologists to view local signals of the heart without explicitly using a mapping catheter, commonly used in minimally-invasive diagnostic or therapeutic procedures.

Several groups have tackled this interesting and challenging problem \cite{gentil2005surface, kachenoura2007surface, kachenoura2008using, kachenoura2009non, kachenoura2009comparison, poree2012surface, mendenhall201012, mendenhall2010implantable}. These works primarily focus on achieving this goal of ECG reconstruction with linear filtering \cite{gentil2005surface, kachenoura2007surface, kachenoura2008using, mendenhall2010implantable}, fixed dipole modeling algorithms \cite{mendenhall201012, mendenhall2010implantable}, and nonlinear reconstruction via a time delay neural network \cite{kachenoura2009non, kachenoura2009comparison, poree2012surface}. The linear approach has been extensively explored in the literature. In \cite{gentil2005surface}, segments of the $12$-lead ECG was synthesized using a single-input and single-output method, meaning a single EGM channel was used to synthesize a single ECG lead. This method was highly dependent on the chosen EGM lead, leading to the logical extension of using all EGM leads for synthesis \cite{kachenoura2007surface, kachenoura2008using}. In this multivariate approach the EGM and the ECG were first both projected into a 3-D space and then three linear filters were calculated between the signals. This provided an indirect way to find the transfer functions between EGM’s and the 12-lead ECG. The synthesis could be performed in the lower dimensional 3-D space and then be back projected to get the full 12-lead ECG.

Following a similar methodology, Mendenhall et al. presented the use of a multivariate linear transfer function \cite{mendenhall201012, mendenhall2010implantable} that directly calculated a transfer matrix between the EGM leads and the 12-lead ECG via penalized linear regression. Although the performance of these linear methods is satisfactory, especially for patients with surface ECG containing only one beat morphology, many improvements are still needed. In a real application, noise, artifacts, and the natural evolution and diversity of the pathology may influence the relationship, over time, between the EGM and the ECG. Thus, the signals exhibit stochastic and nonlinear behavior that cannot be robustly described using the classical linear filtering framework. These same papers also presented the use of a fixed dipole modeling algorithm for ECG reconstruction but could not achieve an average correlation value above 0.5 \cite{mendenhall201012}.

The proposal of a multivariate nonlinear approach was presented in \cite{kachenoura2009non, kachenoura2009comparison, poree2012surface}. This required the simultaneous recording of EGM signals and 12-lead ECG of each patient to train a time-delay artificial neural network (TDNN). This method initially suffered from a difficulty to optimize the architecture structure and hyperparameters \cite{kachenoura2009non, kachenoura2009comparison}, but then received a full analysis and characterization in \cite{poree2012surface}. This method provided the best average correlation results for sinus rhythm heartbeats, but still could not effectively reconstruct diseased morphologies. This method also exhibited the previous multivariate nature, meaning $12$ different TDNN models must be calculated to reconstruct each ECG lead.
 
We contribute to this line of work by proposing a novel architecture named the Nonlinear Regression Convolutional Encoder-Decoder (NRCED). It consists of the cascading of convolutional filtering operations and pointwise nonlinear activations. The closest existing architecture to our work are Deep Autoencoders which are today the state-of-the-art method to denoise, compress, and obtain structured representation of data \cite{eraslan2019single,erhan2010does,tran2017missing, cosentino2020provable}. The NRCED architecture is composed of two parts: a) an encoder that takes its input and maps it to a lower dimensional space, and b) a decoder that takes the output of the encoder and maps it to the output data dimensional space. In particular, the encoder provides the coordinates of the input data in the manifold basis expressed by the decoder. It will be shown that this method provides several improvements over the previously mentioned approaches in terms of reconstruction correlation and generalization. 

The NRCED framework also permits new areas of exploration, such as true multivariate multiple nonlinear regression (using all EGMs as input and outputting the whole 12-lead ECG directly), reconstructing EGMs from the 12-lead ECG, and analysis of the weights used in the model to forego the black box nature of Deep Neural Networks (DNN). Furthermore, we analyze and visualize the weights in the model to obtain new insights into ECG reconstruction. Specifically we use the features learned in the model to form an overcomplete basis of ECG morphologies and leverage those features to create a new diagnostic tool for identifying diseased heartbeats in the patient. We quantified the efficacy of this diagnostic tool by classifying heartbeats of a single patient. This yielded a receiver operating characteristic (ROC) curve with an associated area under the curve (AUC) value of 0.98, demonstrating exceptional discrimination between classes.

Our contribution can be summarized as follows
\begin{itemize}
    \item Proposed a generalized formulation on the problem of multivariate multiple nonlinear regression.
    \item Developed a novel architecture to perform multivariate multiple nonlinear regression.
    \item Provided state-of-the art results for 12-lead ECG reconstruction from a set of EGM leads.
    \item Allowed the reverse mapping of 12-lead ECG to reconstructed EGM.
    \item Created a diagnostic tool for identifying atypical heartbeats from the weights learned in the NRCED model.
\end{itemize}

Section~\ref{sec:NRCED} provides a mathematical description of the problem outlines the formulation of the NRCED along with the exact architecture. Section~\ref{sec:Dataset} outlines the specific dataset being used for this task and Section~\ref{sec:pre-proc} provides the preprocessing specifics to perform this task. Section~\ref{sec:Results} shows the results of the NRCED in practice on our dataset. Section~\ref{sec:Analysis} details the analysis and visualization of the model weights and creation of a new diagnostic tool. Section~\ref{sec:Discussion} and Section~\ref{sec:Conclusion} discuss the results and conclude the work respectively.

\section{Nonlinear Regression Convolutional Encoder-Decoder}
\label{sec:NRCED}
In this work, we consider multi-channel time-series recordings of both the EGM and ECG. We denote by $\left \{ S_n \right \}_{n=1}^{N}$ and $\left \{ X_n \right \}_{n=1}^{N}$ respectively the EGM and ECG samples where $N$ denotes the total number of heartbeats in the dataset. Each heartbeat is a multivariate time-series where the rows define the channel dimension and the columns capture the time dimension. The EGM samples are represented by the following row matrix $S_{n}=\left [ \textbf{s}_n^{(1)},\dots,\textbf{s}_n^{(M_{\text{in}})} \right ]^{\top} \in \mathbb{R}^{M_{\text{in}} \times T}$ where $M_{\text{in}}$ is the number of EGM channels and $T$ the number of time samples per heartbeat, and $\forall i \in \left \{1,\dots,M_{\text{in}} \right \}, \textbf{s}_n^{(i)} \in \mathbb{R}^{T}$ is a column vector. Note that $\mathbb{R}$ is the space of real numbers. Similarly, the ECG samples are expressed by $X_n=\left  [\textbf{x}_n^{(1)},\dots,\textbf{x}_n^{(M_{\text{out}})} \right ]^{\top}$ where $M_{\text{out}}$ is the number of ECG channels and with $\forall i \in \left\{1,\dots,M_{\text{out}}\right \}, \textbf{x}_n^{(i)} \in \mathbb{R}^{T}$.

The problem of multivariate multiple regression lies in the approximation of the function that enables the map from the multivariate input to the multivariate output and can be cast as the following optimization problem
\begin{equation}
    \min_{f \in \mathcal{H}} \sum_{n=1}^{N} \mathcal{L}(f(S_n),X_n),
    \label{eq:opt}
\end{equation}
where $\mathcal{H}$ is a functional space and $\mathcal{L}$ is a loss function. Note that usually $\mathcal{H}$ corresponds to the space of a certain type of parametric function and $\mathcal{L}$ is a reconstruction loss such as the least square error. The result of this optimization problem should be a function that minimizes the loss function on the training data and more importantly that can generalize to unseen test data. In fact in our application this is of critical importance since our aim is to reconstruct a 12-lead ECG with only a new patient's EGM signals as input without the concurrent 12-lead ECG.

This importance of the generalization in our problem motivates us to develop a novel neural network architecture as it has been shown during the last decades that these models are providing the best generalization performances \cite{kawaguchi2017generalization,li2018visualizing,arora2018stronger,geirhos2018generalisation}, and have demonstrated their capability to tackle with super human performance a large number of tasks \cite{ribeiro2020automatic,geirhos2017comparing,mnih2013playing,mnih2015human,tan2019efficientnet}. 

It is now well known that a DNN composed of convolutional operations have an intrinsic bias in their architecture making them a powerful tool to compress, regress, and classify image data \cite{rawat2017deep}. Their application to univariate time-series, in the case of limited amount of data, is usually performed by first obtaining an image representation of the data \cite{hatami2018classification,cosentino2019learnable,balestriero2018spline,zeghidour2019learning}. In this work, we extend this univariate point of view to multivariate inputs by first performing a time-frequency representation of the data, and using this $3$-dimensional tensor as the input of our NRCED. 

\subsection{From Multidimensional Data to Time-Frequency Representation}
Here we propose a formulation of a $3$-dimensional time-frequency representation of the $2$-dimensional multichannel time-series data. As this representation will be performed on both the input and output signal, i.e., $S_n$ and $X_n$, we propose a general formulation. To do so, we define by $Z \in \mathbb{R}^{M \times T}$ a multivariate times-series sample with $M$ channels and $T$ times samples.

It is defined the same way that our data have been previously denoted, that is, $Z = \left [\textbf{z}^{(1)},\dots, \textbf{z}^{(M)} \right ]^{\top}$, where $\forall i \in \left \{1,\dots,M \right \}, \textbf{z}^{(i)} \in \mathbb{R}^{T}$. Note that we formulated our ECG and EGM signals in the exact same manner, but have not specified the number of channels $M$ here as to allow generalization. We define the operator $\mathcal{W}$ that takes as input a multivariate time-series sample and outputs its $3$-dimensional time-frequency representation as
\begin{equation}
\label{eq:_tensor_rep}
    \mathcal{W}(Z) = \left [ W^{(1)}(Z),\dots, W^{(M)}(Z) \right ],
\end{equation}
where $\forall i \in \left \{1,\dots,M \right\}$
\begin{equation}
     W^{(i)}(Z)= \left [  \textbf{z}^{(i)} \circ \textbf{h}_{1},\dots, \textbf{z}^{(i)} \circ \textbf{h}_{K}                \right ]^{\top},
\label{eq:output_conv_rep}
\end{equation}
where $\forall k \in \left \{1,\dots,K \right \}, \textbf{h}_k \in \mathbb{C}^{T}$ defines the time-frequency filter in complex space and $\circ$ is the convolution operator. Thus we have that $ W^{(i)}(Z)$ is a complex matrix with frequency information of the signal $\textbf{z}^{(i)}$ in its row dimension and time information in its columns. Now the concatenation of these $2$-dimensional time-frequency representations leads to the $3$-dimensional representation of the signal $Z$ given by $\mathcal{W}(Z) \in \mathbb{C}^{M \times K \times T}$. For our data, we show in Section~\ref{sec:pre-proc} that we use as the filters $\textbf{h}_k$ windowed Fourier filters thus making $\mathcal{W}$ a $3$-dimensional short-time Fourier transform operator applied to multivariate time-series. In the same fashion, other time-frequency or any image representations of the data , e.g., the wavelet transform, can also be used in our framework.

\subsection{NRCED Architecture}
We now consider the nonlinear regression framework having as input and output data the $3$-dimensional time-frequency representation respectively defined in the time-frequency domain by $\left \{\mathcal{W}(S_n) \right \}_{n=1}^{N} \in \mathbb{C}^{M_{in} \times K \times T}$ and $\left \{\mathcal{W}(X_n) \right \}_{n=1}^{N} \in \mathbb{C}^{M_{out} \times K \times T}$ by applying the procedures outlined in (\ref{eq:_tensor_rep}) and (\ref{eq:output_conv_rep}) on our individual heartbeats $X_n$ and $S_n$ for all $N$.

Given these input and output training data, we aim at providing a parametrized functional space $\mathcal{H}$ that is designed specifically for our problem and that has the highest generalization capability. Inspired by Encoder-Decoder architectures that allow the approximation of the data manifold, we propose the NRCED architecture. That is, a cascade of convolutional transformations and nonlinearities with a bottle-neck dimension aiming at approximating the dimension of the underlying manifold.
Our architecture is first composed of an encoder, $\textbf{E}$ taking as input the $3$-dimensional time-frequency representation $\mathcal{W}(S_n)$ and outputting a low dimensional latent representation of our data $\textbf{E}(\mathcal{W}(S_n);\theta_{\text{e}})$, where $\theta_{\text{e}}$ is the set of all learnable parameters in the encoder. The second half of the architecture contains a decoder $\textbf{D}$ taking as input the output of the encoder, $\textbf{E}(\mathcal{W}(S_n);\theta_{\text{e}})$, and giving as output an approximation of $X_n$, $\textbf{D}( \textbf{E}(\mathcal{W}(S_n));\theta_{\text{d}})$, where $\theta_{\text{d}}$ is the set of all learnable parameters in the decoder. In fact, both $\textbf{E}$ and $\textbf{D}$ are nonlinear functions that are composed of layers of convolutions and pointwise nonlinearities. Each of them have their own internal parameters that are the convolutional filters and fully connected layer parameters. These are the parameters that one needs to learn as to best approximate $\mathcal{W}(X_n)$, that is, our optimization problem can now be expressed as

\begin{equation}
    \min_{\theta_{\text{e}}, \theta_{\text{d}}} \sum_{n=1}^{N} \mathcal{L}(\textbf{D}(\textbf{E}(\mathcal{W}(S_n))),\mathcal{W}(X_n)),
\end{equation}
where we observe that the space of optimization defined in (\ref{eq:opt}) has been parametrized by this specific architecture. 

Now, there is an intrinsic ability of this architecture to extract the underlying structure of the data. In the same fashion as Encoder-Decoders do, the dimension of the output of the encoder is smaller than both the dimension of the input data and output data  which leads to the encoding of the salient features in the data \cite{goodfellow2016deep}. In particular, it can be mathematically proven that the encoder transforms the input data as to obtain their coordinate in the manifold basis given by the decoder \cite{lei2020geometric}. In the same fashion that Principle Component Analysis (PCA) finds the coordinate of the data in the optimal orthogonal basis that minimizes the reconstruction of the data under a least-square error \cite{abdi2010principal}, our architecture aims at finding the coordinates of the EGM data in the optimal basis that minimizes the loss function $\mathcal{L}$. As a matter of fact, it is realistic to consider that there is an underlying manifold of the data unifying both the EGM and ECG as they represent both the same underlying physical process, that is, the polarization and depolarization of the myocardium.

\begin{figure*}
\centerline{\includegraphics[width=\textwidth,height=\textheight,keepaspectratio]{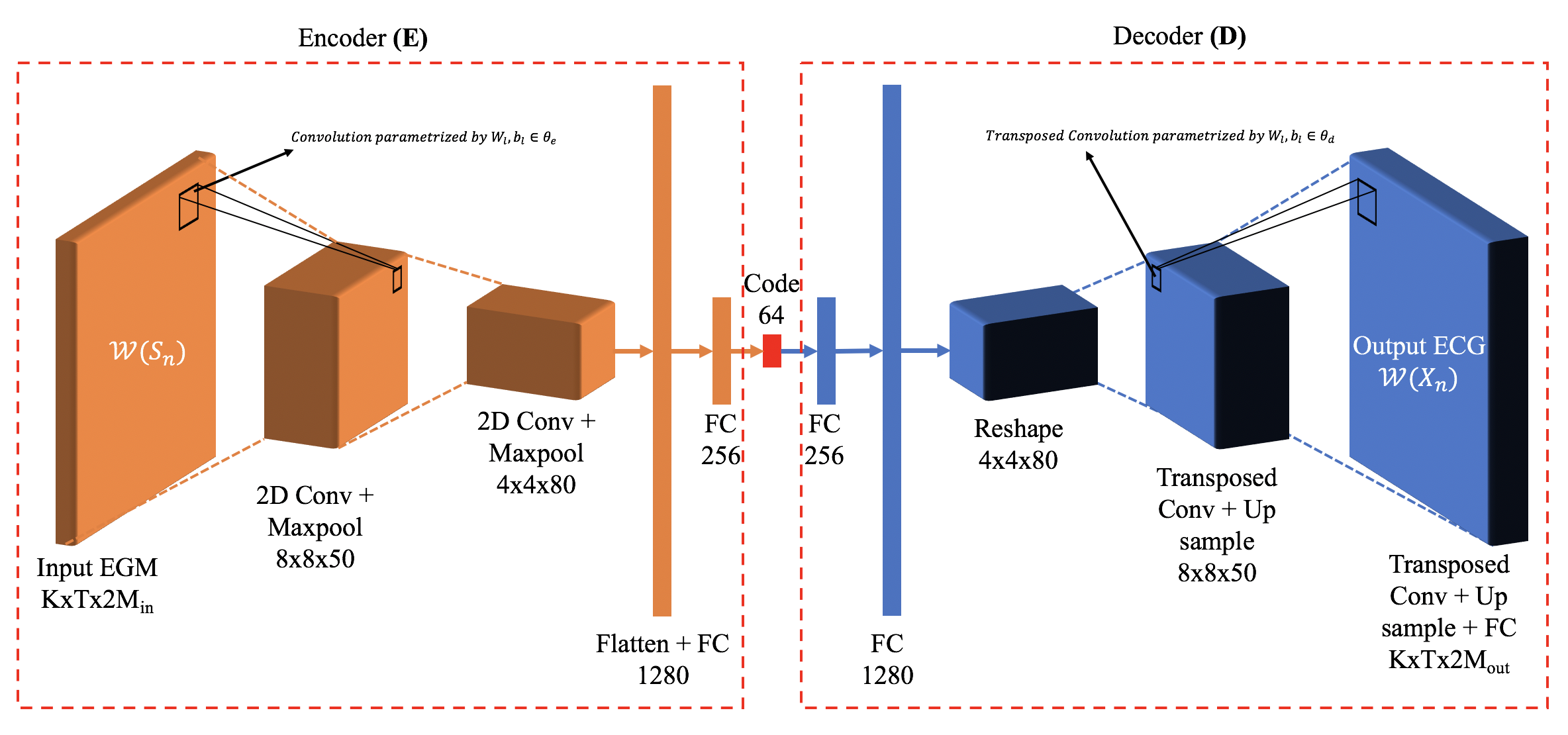}}
\caption{Nonlinear Regression Convolutional Encoder-Decoder model (FC means a fully connected layer). This is a modified convolutional Encoder-Decoder setup, with two convolution/ transposed convolution layers and six fully connected layers. The activation function used was the hyperbolic tangent function and the learning rate was $5*10^{-3}$. The novelty in this architecture comes from the mismatch in dimensions between the input and output.}
\label{fig: Architecture}
\end{figure*}

The network utilizes convolution, max pooling, and fully connected layers in the encoder portion of the network. The decoder portion of the network utilizes upsampling layers, transposed convolution layers, and fully connected layers to produce the output. More details regarding these common layers can be found in \cite{goodfellow2016deep}. Note that the convolution, transposed convolution, and fully connected layers require the learning of specific weights $W_l$ and biases matrices $b_l$ to perform the operations. Thus we can say $\theta_{e} = \left \{W_1,\dots,W_l, b_1,\dots,b_l \right \}$ and $\theta_{d} = \left \{W_{l+1},\dots,W_L, b_{l+1},\dots,b_L \right \}$, where $l$ is the layer number of the network. The size of the filters in the convolution and transposed convolution layers were picked in a hyperparameter search. Every layer also contains batch normalization and dropout to improve generalization accuracy. A visualization of the network used is shown in Fig. \ref{fig: Architecture}.

Once we have reconstructed an estimated 12-lead ECG $\mathcal{W}(X_n)$ we must use an appropriate loss function $\mathcal{L}$ to inform the network how to train. A common loss function in a regression application like ours is to calculate the mean squared error loss between our reconstruction and the actual corresponding ECG. However, the mean squared error loss focuses on replicating the exact values between two signals, but we are more interested in replicating the proper morphologies between two signals since the information in an ECG is contained in the shape of the waveform. To achieve this we use the Pearson correlation coefficient \cite{benesty2009pearson}. This metric focuses on capturing the morphology of a waveform and helps to reduce the reconstruction of noise in our output signal \cite{benesty2008importance}. We can calculate the Pearson correlation coefficient between two vectors as
\begin{equation}
\rho(\textbf{a}, \textbf{b}) = \frac{\mathbb{E}[(\textbf{a}-\mathbb{E}[\textbf{a}]) (\textbf{b}-\mathbb{E}[\textbf{b}])^{\top}]}{\mathbb{E}[\textbf{a}\textbf{a}^{\top}] \mathbb{E}[\textbf{b}\textbf{b}^{\top}]},
\label{eq: rho}
\end{equation}
where $\textbf{a}$ and $\textbf{b}$ are two arbitrary vectors of equal length, and E is the expectation operator. The Pearson correlation is a statistic that measures linear correlation between two variables. It has a value between $+1$ and $-1$. A value of $+1$ is total positive linear correlation, $0$ is no linear correlation, and $-1$ is total negative linear correlation \cite{benesty2009pearson}.

We can use the Pearson correlation coefficient to write our objective function for the model, but we must first make the data compatible with (\ref{eq: rho}). We can rewrite $\textbf{D}(\textbf{E}(\mathcal{W}(S_n))$ as $\hat{\mathcal{W}}(X_n)$ where this represents the output of the model, which is an estimation of our $12$-lead ECG $\mathcal{W}(X_n)$ in the time-frequency domain. Next, we denote with an underline the operator taking as input a tensor and reshaping it as a vector. We then express the loss function as
\begin{equation}
    \min_{\theta_{\text{e}}, \theta_{\text{d}}} -\sum_{n=1}^{N} \rho(\underline{\hat{\mathcal{W}}}(X_n),\underline{\mathcal{W}}(X_n)),
\label{eq: loss}
\end{equation}
where we have specified the exact loss function $\mathcal{L}$ as the Pearson correlation coefficient defined in (\ref{eq: rho}). Note that there is a negative sign in (\ref{eq: loss}) because gradient descent minimizes the objective function but we want to maximize the Pearson correlation coefficient (bring as close to 1 as possible) between $\mathcal{W}(X_n)$ and our reconstruction. The Pearson correlation coefficient is what is used for our model's loss function and our metric for accuracy of reconstruction later in the paper.

In practical applications we do not calculate this per sample, but instead in batches of samples to be more efficient and to ensure proper convergence of the model \cite{bengio2012practical}. This loss is used to update the parameters via the Adam optimizer. The Adam optimizer allows an efficient implementation of stochastic gradient descent to minimize our objective function in (\ref{eq: loss}) \cite{kingma2014adam}.

\section{Description of the Dataset}
\label{sec:Dataset}
A dataset collected from 14 patients was used to evaluate the performance of the NRCED model for ECG reconstruction. The ECG and EGM signals were recorded simultaneously during a cardiac ablation procedure. Each record of the database is composed of
\begin{itemize}
\item Twelve standard surface ECG channels. Namely leads I, II, III, aVR, aVL, aVF, and V1:V6.
\item Five EGM channels measured by electrodes on a catheter placed inside the Coronary Sinus. 
\end{itemize}
The data was obtained from patients undergoing cardiac ablation procedures. During these procedures, it is routine to record surface ECG and EGM via mapping catheter. For each patient EGM was obtained from the coronary sinus. By virtue of the procedure, the recordings are of different lengths per patient and contain a mix of sinus rhythm and diseased heartbeats. This provides a diverse dataset for the NRCED model, but also adds complexity to the reconstruction. This also means that each patient contributes a different number of heartbeats in their dataset $N$. Each patient file was first randomly shuffled then segmented in half. The first half of concurrent ECG and EGM signals is used during the training step and the second half is devoted to testing performance evaluation. The data is summarized in Table 1. 

\begin{table}[htbp]
\caption{Summary of Dataset}
\begin{center}
\begin{tabular}{|c|c|}
\hline
\textbf{Patient}&\multicolumn{1}{|c|}{\textbf{Number of Heartbeats (N)}} \\
\hline
2 & 4765 \\ 
\hline
3 & 2309 \\ 
\hline
4 & 401 \\  
\hline
5 & 1752 \\
\hline
7 & 3934 \\ 
\hline
8 & 3017 \\ 
\hline
9 & 2593 \\  
\hline
13 & 6635 \\  
\hline
17 & 3102 \\
\hline
18 & 2326 \\  
\hline
19 & 5497 \\  
\hline
24 & 1591 \\
\hline
25 & 1827 \\ 
\hline
26 & 2917 \\  
\hline
Total & 39868 \\  
\hline
\end{tabular}
\label{tab1}
\end{center}
\end{table}

\section{Pre-Processing}
\label{sec:pre-proc}
The ECG and EGM signals were recorded simultaneously during a cardiac ablation procedure. The signals were initially obtained at a sampling frequency of 1000 Hz. The data were subsequently bandpass filtered with a 5th order Butterworth filter with cutoff frequencies of $3$ and $50$ Hz. A cutoff of 3 Hz eliminates baseline wander and a cutoff of 50 Hz eliminates powerline interference, electromyographic noise, and electrode motion artifact noise \cite{kher2019signal}. Normal filtering is done only in the forward direction in time. This causes a time delay and thus a change in the phase of the signal. We can easily fix that by also filtering the signal backwards in time \cite{kormylo1974two}. This is called a zero phase filter and is beneficial here since we want to preserve all information in time without introducing distortions.

\begin{figure}
\centerline{\includegraphics[width=\columnwidth,keepaspectratio]{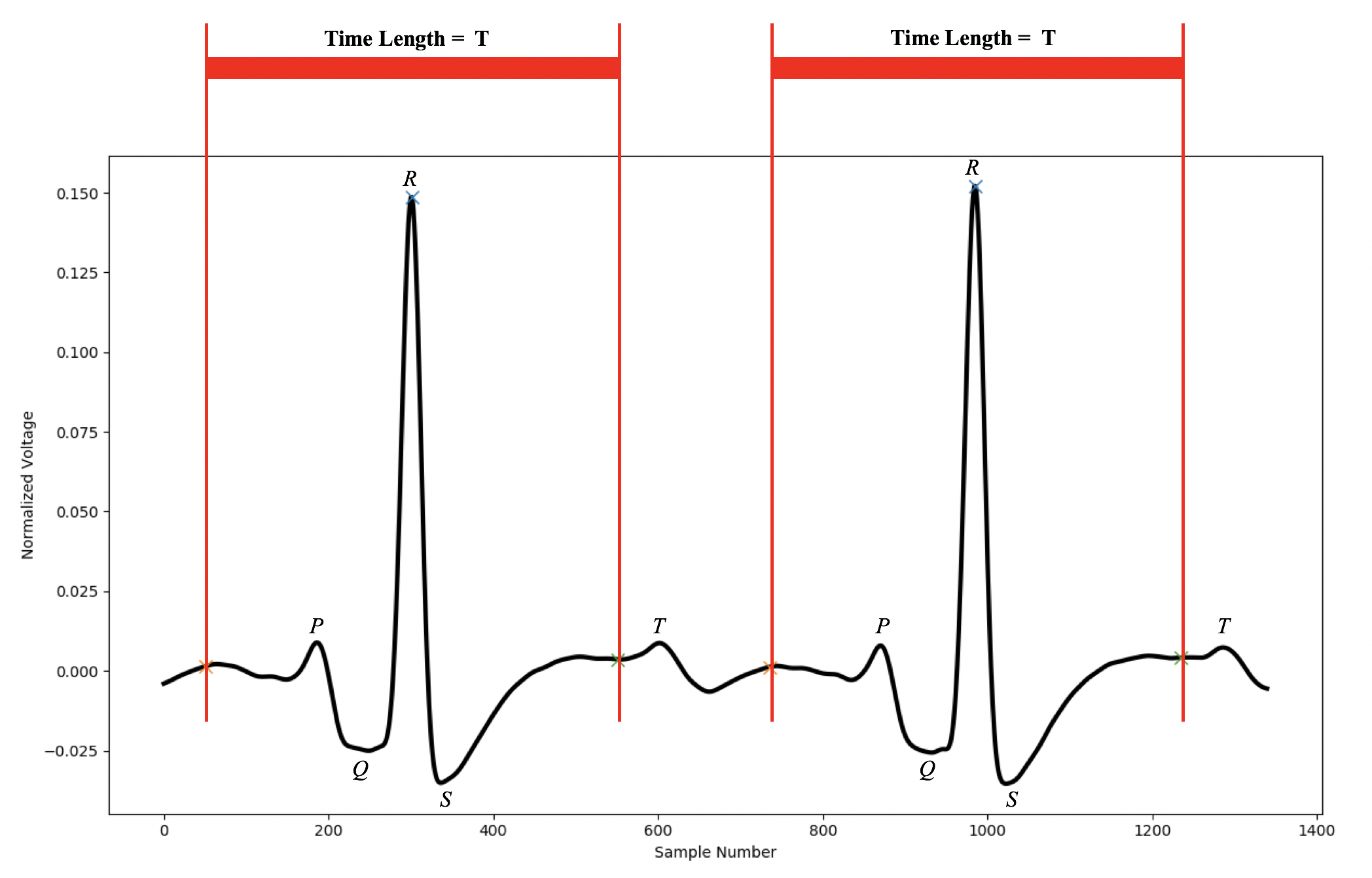}}
\caption{Segmentation procedure of ECG and EGM waveforms. First the R wave of the ECG is found using a peak detection algorithm (blue x on the figure). We then extract blocks of data around each R wave peak measuring 250 samples (denoted as orange and green x). This choice of length encompasses the whole QRS complex and P wave of the signal and is denoted $T$ as from our mathematical formulation. We extract our snippets in this manner because the heartbeats in our data set are dynamic and change in frequency throughout each recording. Thus it is crucial to have an interval of time between snippets to ensure that we do not have overlapping blocks of data.}
\label{fig: blocks}
\end{figure}

Next, we perform segmentation on the data. To explain this process clearly we must introduce some common nomenclature associated with $12$-lead ECG waveforms. A heartbeat shown on an ECG has five important features denoted the P, Q, R, S, T waves \cite{cardiovascularbook}. These can be seen labeled in Fig. \ref{fig: blocks}. Each wave is indicative of a specific physiological process of the heart during a heartbeat. It's common to talk about the Q, R, and S wave together since they occur in succession and all relate to the contraction of the heart's ventricles. In this scenario we denote this portion of the signal as the QRS complex \cite{cardiovascularbook}. The QRS complex provides a wealth of information about the synchronization of the heart's ventricles and has been shown to be a strong biomarker for overall cardiac health by indicating the presence of premature ventricular complexes \cite{moulton1990premature}. Due to this property of the QRS complex, we will focus primarily on extracting this portion of the ECG signal. 

To segment our ECG and EGM we take the full filtered ECG and find the R wave peak for every heartbeat. This is achieved using a package in python called numpy that has a peak finding built-in function. We then use the peak locations to extract snippets of the ECG and EGM waveforms around that peak location. Each snippet is centered at the R wave peak location and extends $250$ samples in both directions. The length of the snippet was chosen to encompass the QRS complex and P wave. Note that since our sampling rate is $1000$ Hz, $250$ samples amounts to $0.25s$ of data in each direction from the R wave. This provides a large enough window to encapsulate a standard QRS complex ($0.08s-0.11s$) or even a prolonged QRS complex($>0.11s$), and the P wave (the time between the P and R wave is between $0.12s$ and $0.22s$) \cite{mason2007electrocardiographic}. We can see in Fig. \ref{fig: blocks} that we are indeed encompassing the QRS complex and P wave in our extracted snippets of data.

Each snippet is then down sampled by a factor of 2 in order to decrease the dimension of the data. A reduced sampling frequency of 500 Hz will still capture all relevant frequency components of our ECG \cite{kwon2018electrocardiogram}. This gives our specific value for $T$, the number of time samples, as $250$. Next we center the data by subtracting the mean from each snippet and then normalizing each snippet. These steps are commonplace in neural network applications in order to avoid exploding or vanishing gradients \cite{ioffe2015batch}. These steps allow us to obtain our processed data $\left \{ S_n \right \}_{n=1}^{N}$ and $\left \{ X_n \right \}_{n=1}^{N}$ described in section \ref{sec:NRCED}.

As mentioned in section \ref{sec:NRCED}, for the data to be compatible with a convolutional Encoder-Decoder architecture the input data is required to be the type of an image. Researchers in the speech recognition and audio processing domain have addressed this problem by leveraging time-frequency representations to convert multivariate time series into multichannel images \cite{amodei2016deep, parveen2004speech, xu2014regression}. We can emulate these prior approaches by applying (\ref{eq:output_conv_rep}) to the individual time series in $\left \{ S_n \right \}_{n=1}^{N}$ and $\left \{ X_n \right \}_{n=1}^{N}$. We choose $\textbf{h}_k$ to be the basis of windowed Fourier filters thus performing the Short-Time Fourier Transform (STFT) on our processed samples. The STFT defines a particularly useful class of time-frequency distributions which specify complex amplitude versus time and frequency for any signal \cite{huang2019ecg}. This transformation is also invertible, allowing us to move easily between the time domain and the time-frequency domain \cite{oppenheim2001discrete}. We can specify our filters  $\textbf{h}_k$ as
\begin{equation}
\textbf{h}_k=w(t-t^*)e^{-j k t},
\end{equation}
where $w(t)$ is a fixed length window function, $t$ is the time sample, $k$ is the frequency, and $t^*$ is the time point of the total length $T^*$ window function. In this proposed method we adopt the Hann window, whose definition is given below
\begin{equation}
w(t)=\begin{cases} 0.5\left [{ {1-\cos \left ({{\dfrac {2\pi t}{T^*}} }\right)} }\right],&0\le t\le T^* \\ 0,&otherwise \\ \end{cases}
\end{equation}

One last nuance in processing is that each 3-dimensional time-frequency representation $\mathcal{W}(X_n)$ and $\mathcal{W}(S_n)$ is complex valued. This poses a complication since operating on complex values with neural networks can be difficult, but we must preserve the real and imaginary portions of the spectrograms so that we can later reconstruct our signals into the time domain. We navigate this issue by stacking $\text{Re}\{\mathcal{W}(X_n)\}$ and $\text{Im}\{\mathcal{W}(X_n)\}$, thus doubling the number of channels of our signals. Our final processed signals are given by $\mathcal{W}(X_n) \in \mathbb{R}^{2M_{in} \times K \times T}$ and $\mathcal{W}(S_n) \in \mathbb{R}^{2M_{out} \times K \times T}$ where $K$ and $T$ are the same variables from (\ref{eq:output_conv_rep}) representing frequency and time respectively and are both $16$ to produce a $16 \times 16$ pixel image. This approach to convert waveforms into STFT images has been widely used in the speech recognition and audio processing domain, with \cite{amodei2016deep} using the full STFT to reduce the dimensionality of the raw audio waveforms and \cite{parveen2004speech, xu2014regression} using the magnitude of the STFT as the input data.

\begin{figure*}
\centerline{\includegraphics[width=\textwidth,height=\textheight,keepaspectratio]{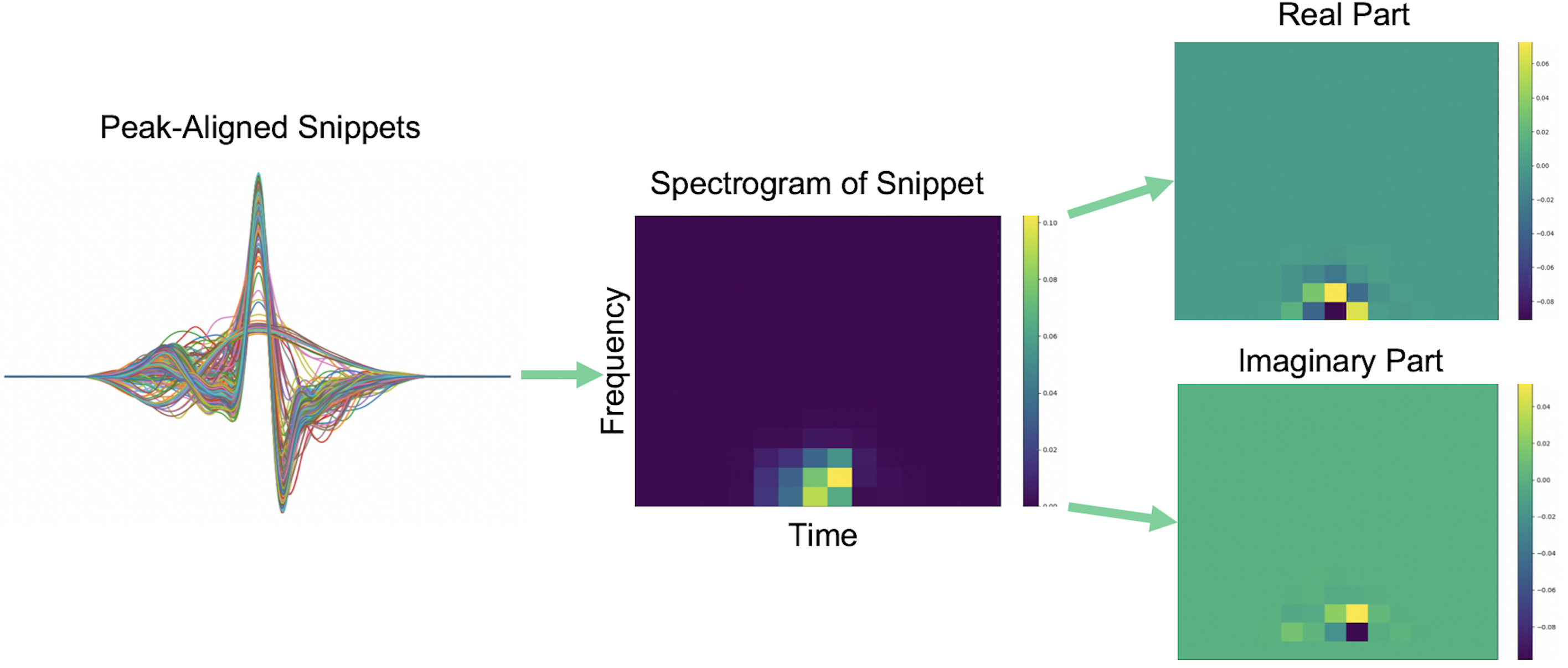}}
\caption{Pre-processing pipeline of the ECG and EGM signals.
Left image is all processed heartbeats for one patient, middle image is the absolute value of the STFT of our snippets, and the right image is the splitting of the STFT into it's real and imaginary portion. This process doubles the number of channels of our data.
}
\label{fig: complex}
\end{figure*}

This conversion into the time-frequency domain is summarized in Fig. \ref{fig: complex}. Conversion into images does truncate the waveform, but the most relevant information to cardiologists is stored in the QRS complex of the ECG signals, which is maintained in each snippet. The heartbeats can also be reconstructed as a sequence in time so that cardiologists can still observe changes in waveform morphology over time. Thus, the change from continuous waveforms to discrete images does not discard much relevant information.

\section{Results}
\label{sec:Results}
With the model and pre-processing pipeline described we then tested the algorithm in several different settings. First, we tested reconstruction of the model in a patient-specific manner, meaning we split each patient’s dataset in half for training and testing and trained the model on each individual patient. Subsequently we constrained the model by only allowing one input EGM signal, thus allowing us to see how dependent the model’s reconstruction is on the number of input EGM signals. Next we pooled all patients but one as the training set and tested on an entirely new patient to gauge the generalizability of the model. Lastly, we trained the algorithm in the reverse direction to see if we could reconstruct EGM signals from a $12$-lead ECG.

\subsection{Patient-Specific Reconstruction}
As mentioned above, we have a dataset of 14 patients containing sinus and diseased heartbeats. We will split each data set into a test and train set at a 50-50 ratio and measure the reconstruction correlation coefficient as in \ref{eq: rho} per patient. In order to measure reconstruction correlation, we do a feedforward run through the model with the test set and save it. The output of the model will be in the time-frequency domain so we must use the inverse short time Fourier transform to get our original time-series waveforms. Once the signals are back in the time domain, we calculate the Pearson correlation coefficient between the reconstructed output and the actual ECG signals. Note that we are calculating this in the time domain because this is the signal a cardiologist would actually look at in practice.

\begin{figure*}
\centerline{\includegraphics[width=\textwidth,height=\textheight,keepaspectratio]{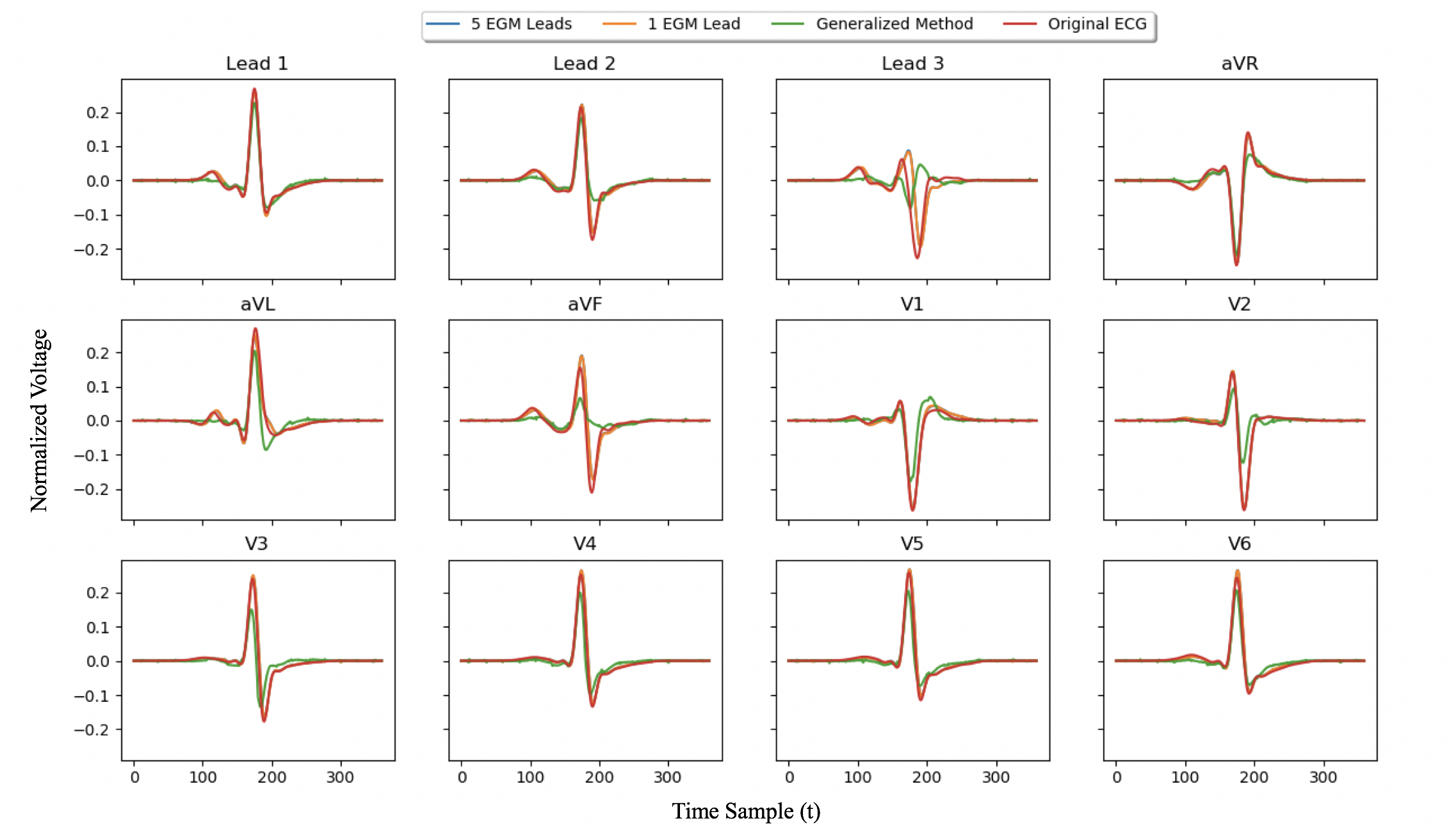}}
\caption{One ECG heartbeat sample for patient 5. The x axis is the time sample and the y axis is the normalized voltage of the waveforms. The red trace shows the actual ECG of the patient while the other three traces show our calculated reconstructions with different methods. The blue trace is our reconstruction when we use all five input EGM leads into the NRCED model. We see a convincing overlap in the two signals for all leads indicating an accurate reconstruction of our 12-lead ECG from a set of EGMs. The orange trace is our reconstruction when using only one input EGM lead into the model. Even with only one input lead to our model, we get a similar quality reconstruction as with all five leads as input. This provides confidence that this method would work with input leads positioned where a pacemaker typically has them. Lastly, the green trace is our reconstructed ECG when trained on 13 patients and tested on one. As compared to the previous reconstructed traces, we see a lower reconstruction accuracy. This is to be expected since the model never sees the patient's data and thus must rely on all the other patient's data to reconstruct. However, the traces still show proper directions of deflection in most cases.}
\label{fig: forwardreconstruction}
\end{figure*}

\begin{figure}
\centerline{\includegraphics[width=\columnwidth, keepaspectratio, trim=4 4 4 4, clip]{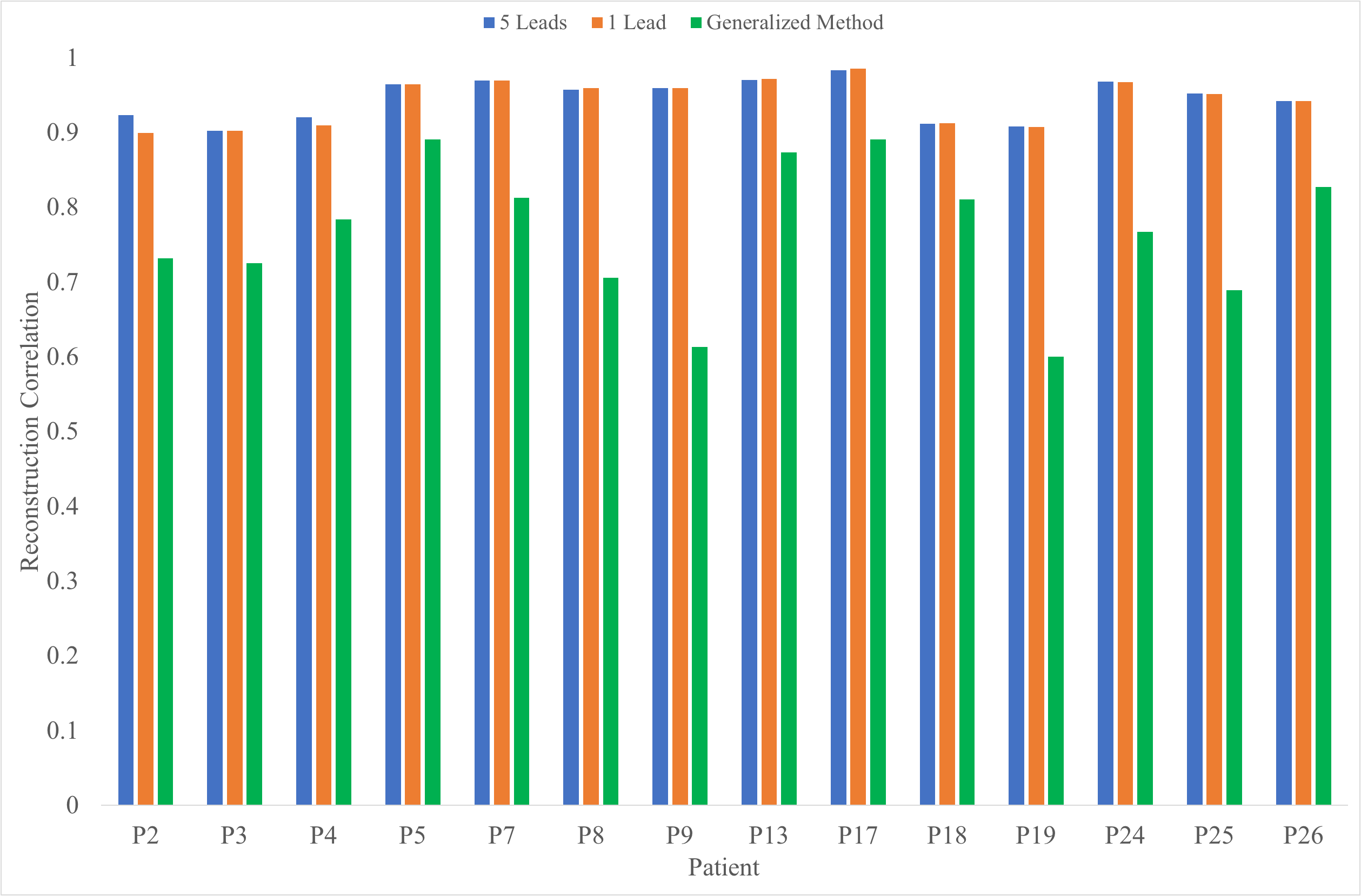}}
\caption{Average correlation between reconstructed and original ECG for all channels. This is calculated per patient by using (\ref{eq: rho}) and then taking the average over $n$. The colors used match the colors used in Fig. \ref{fig: quantfeedforwardreconstruct} to maintain clarity. The blue bars are the correlation values when using all five input EGM leads. The correlation value never dipped below $0.9$ for all patients in this reconstruction scheme. The orange bars are the correlation values when using only one input EGM lead. Every patient exhibits no significant loss in reconstruction correlation. The largest drop in correlation is for patient 2 with a drop of $2\%$. Interestingly we see a slight increase in correlation for patients $13$, $17$, and $18$. Lastly, the green bar indicates the correlation between the reconstructed ECG and original for each patient when trained on all other patients. The correlations are highly dependent on the patient of interest since each patient will exhibit different morphologies in their 12-lead ECG. The highest correlation value of $0.89$ was achieved with patient $5$ and $17$ and the lowest correlation value of $0.60$ was produced by patient $19$. This indicates that we require more data to achieve a perfectly generalized implementation of our algorithm.}
\label{fig: quantfeedforwardreconstruct}
\end{figure}

The original ECG and an example reconstruction using all five EGM leads is shown in Fig. \ref{fig: forwardreconstruction}. We see that for every lead except lead $3$ that the reconstruction is nearly perfect as indicated by the strong overlap between the red and blue trace in Fig. \ref{fig: forwardreconstruction}. To quantify how accurate our reconstructions were we calculated the Pearson correlation coefficient between every reconstructed and actual heartbeat and then took the average value. This was done for every patient in the dataset. The values obtained are shown by the blue bars in Fig. \ref{fig: quantfeedforwardreconstruct}. 

The correlation values range from $0.90$ to $0.98$ which is a major improvement over the state of the art method \cite{poree2012surface}, considering we are predicting 100 times more heartbeats and are reconstructing sinus and diseased heartbeats. We also see that the reconstruction correlation varies between patients, highlighting the differences that exist between each patient's data. This can be due to differing physiology of the heart, different stages of diseased hearts, and the differing number of peaks per patient which can lead to more challenging training conditions. 

\subsection{Reconstruction with One EGM Lead}
One limitation to our above results is that we used five EGM signals from the coronary sinus to reconstruct, while the most advanced pacemakers measure from only $2$-$3$ sites in the heart. Although we can not change the location of our input EGM signals, we can attempt to reconstruct with less leads to better simulate the conditions of a clinical setting. Reconstruction was attempted using only one input EGM for each patient. The chosen EGM lead is the most challenging lead to reconstruct with since it is predominately sensing atrial signals instead of a balanced mix of atrial and ventricular signals. To do this experiment, the model was retrained using only the first EGM channel in our dataset for each patient.

The orange traces in Fig. \ref{fig: forwardreconstruction} show the ECG reconstruction when only using one input EGM lead. The reconstructions closely resemble the reconstructed traces using all five EGM leads as input, as evidenced by the orange and red traces in Fig. \ref{fig: forwardreconstruction}. We quantified the average correlations for this method in the orange bars of Fig. \ref{fig: quantfeedforwardreconstruct} and indeed see that the correlation values do not change significantly for all patients in the dataset. 

\subsection{Generalizability}
The ideal scenario to use this algorithm is in a non-patient-specific manner so that we don’t have to gather training data for each patient. To test the ability of the algorithm to generalize, it was trained on all patients but one and then tested on the remaining patient. This procedure was repeated for every patient.

In the green traces of Fig. \ref{fig: forwardreconstruction} we see the same reconstructed beat for patient $5$ when the model is trained on all other $13$ patients. This trace shows the most deviation from our original ECG trace in red. This is to be expected since the model never trains on any physiological data from patient $5$, but is still tasked with reconstructing patient 5's data. The quantification of the correlation between the green and red traces is shown in the green bars of Fig. \ref{fig: quantfeedforwardreconstruct}. We see that the correlation varies heavily between patients. The highest correlation value of $0.89$ was achieved with patient $5$ and $17$ and the lowest correlation value of $0.60$ was produced by patient $19$. This indicates that a larger dataset would be needed to use the algorithm in a generalized fashion with high confidence in the reconstructions.

\subsection{Reverse Reconstruction}
In some scenarios, converting from the ECG to a local electrogram provides an immense benefit. One example is during ablation studies. A cardiac ablation is a procedure to scar or destroy tissue in your heart that's allowing erroneous electrical signals to cause an abnormal heart rhythm. The biggest difficulty in this procedure is locating the source of irregularity on the heart to ablate. The surgeon starts by reading the $12$-lead ECG to narrow the search region of the irregularity. They then have to map the heart locally, obtaining EGM signals with an electrode attached to a catheter. In this scenario we could do the reverse reconstruction to show what a local EGM signal would look like in this region and further narrow the region of interest without needing to map the heart electronically in an invasive manner. 

To do this reverse reconstruction we made the input be the 12-lead ECG and the desired output be the five EGM signals. The model is the same except the feedforward path was the backward pass in the normal reconstruction problem. To make this clear we can define our objective mathematically below as
\begin{equation}
    \min_{\theta_{\text{d}}, \theta_{\text{e}}} -\sum_{n=1}^{N} \rho(\hat{\underline{\mathcal{W}}}(S_n),\underline{\mathcal{W}}(S_n)),
\end{equation}
where $\theta_{e}$ and $\theta_{d}$ are the parameters of our model when trained in the reverse direction and $\hat{\underline{\mathcal{W}}}(S_n)$ is the flattened output of the model which can also be expressed in tensor form as $\textbf{E}(\textbf{D}(\mathcal{W}(X_n)))$. Note that the position of $\textbf{E}$ and $\textbf{D}$ is switched due to the reversal of the algorithm. This formulation is identical to (\ref{eq: loss}) except that we have replaced $X_n$ with $S_n$ so that we may reconstruct the local electrogram signals.

\begin{figure}
\centerline{\includegraphics[width=\columnwidth,keepaspectratio]{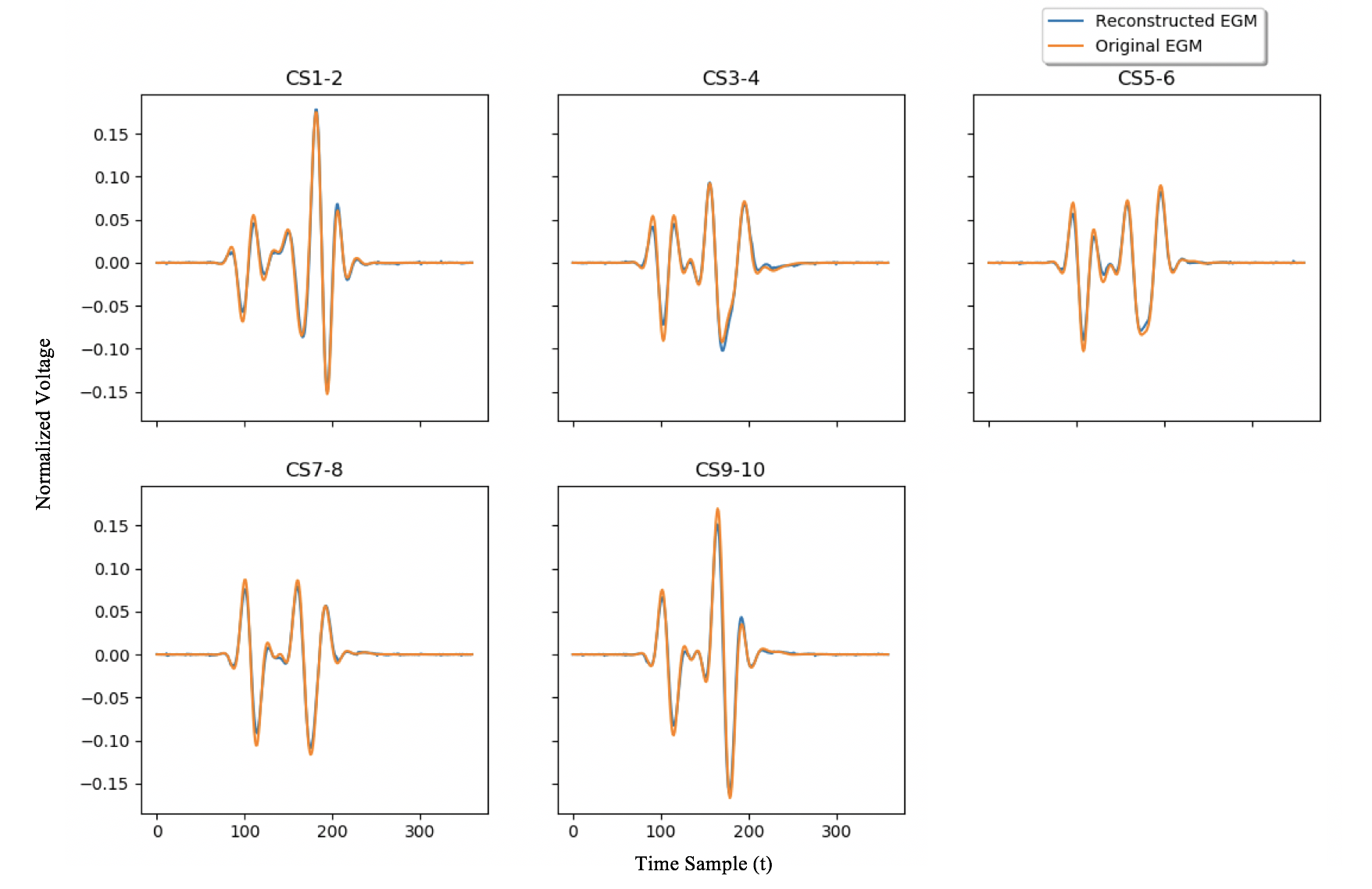}}
\caption{Sample reconstruction of EGM signals for patient $24$. The x axis is the time sample and the y axis is the normalized voltage of the waveforms. For this patient we see a nearly flawless reconstruction of the EGMs. We also see that the intracardiac EGMs have much more complicated morphologies than the 12-lead ECG in general.
}
\label{fig: reverse}
\end{figure}

\begin{figure}
\centerline{\includegraphics[width=\columnwidth,height=0.3\textheight, trim=4 4 4 4, clip]{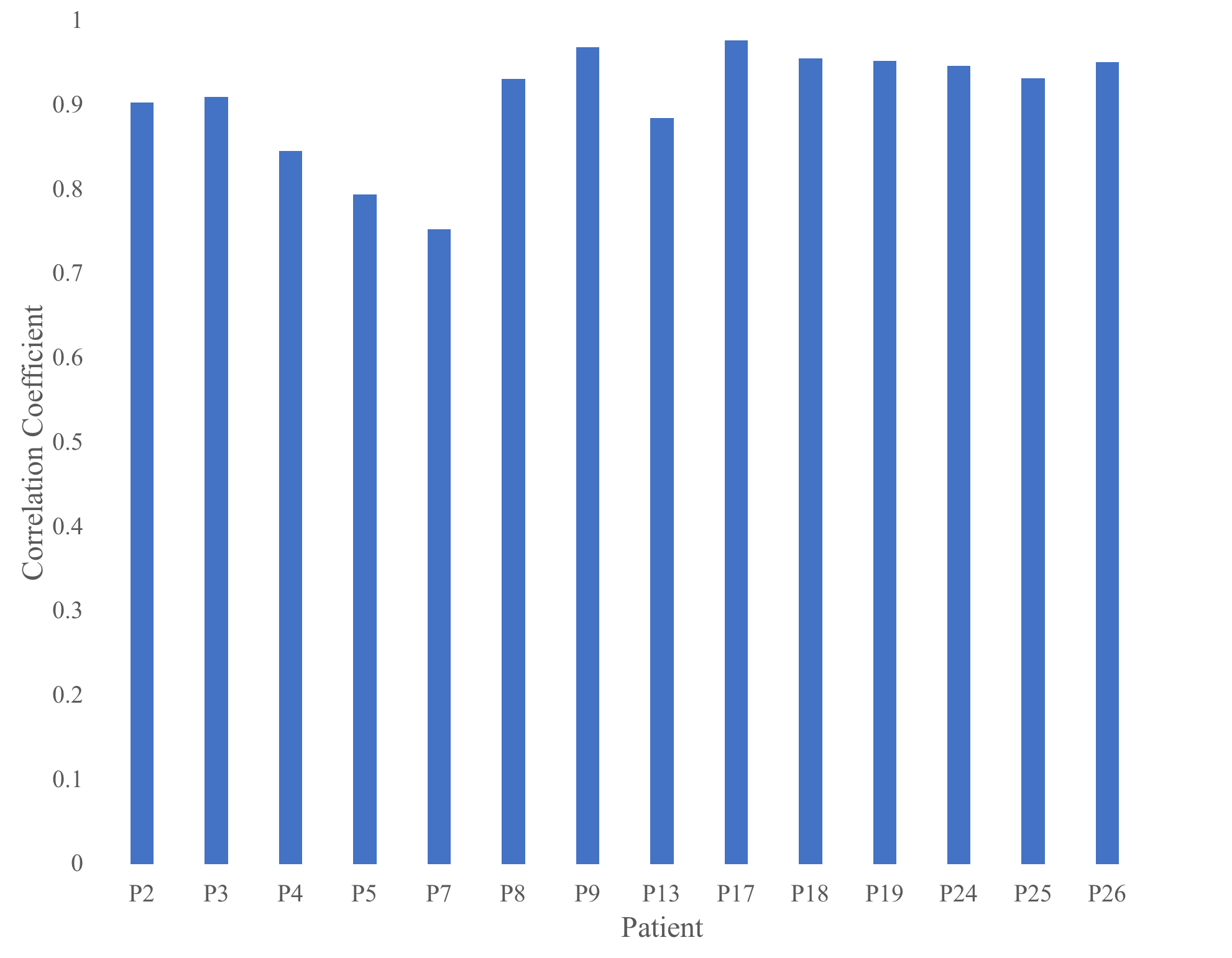}}
\caption{Correlation between reconstructed and original EGM for all patients. Most patients exhibited a reconstruction correlation above 0.9, thus implying that the model works effectively in either direction.
}
\label{fig: reversecorr}
\end{figure}

An example reconstruction is shown in Fig. \ref{fig: reverse} and a quantification of the results is shown in Fig. \ref{fig: reversecorr}. We see in Fig. \ref{fig: reversecorr} that we achieved a correlation coefficient between $0.75$ and $0.97$ depending on the patient, but a majority of the patients had correlations above $0.9$. This showed that we can indeed run the model in the reverse direction and obtain local signals of the heart without invasively measuring the signals directly as long as we have the $12$-lead ECG available.

\section{Conceptual Analysis}
\label{sec:Analysis}
Although neural networks provide a great way to learn nonlinear functions between variables, they achieve this in a black-box manner \cite{zhang2018opening, heckerling2003entering}. The function is contained in thousands or millions of parameters fine tuned by gradient descent and as we increase the number of layers in the model, interpretability becomes incredibly tedious. However, work has been done to interpret the weights in neural networks \cite{zhang2018opening, heckerling2003entering, zeiler2014visualizing}. Following suit, we will visualize and analyze the weights in our model to reveal new insights into ECG reconstruction and utilize the weights to create a new diagnostic tool for detecting atypical heartbeats.

The features lying in the data space that are easily interpretable are stored in the final weight matrix $W_L$. We can represent this as
\begin{equation}
\hat{\underline{\mathcal{W}}}(X_n) = W_L\textbf{G}(\mathcal{W}(S_n))).
\label{eq: last layer}
\end{equation}
Note that we are using the feedforward version of the model for the whole analysis section. From before (\ref{eq:_tensor_rep}), $\mathcal{W}(S_n)$ is our input EGMs that have been transformed into the time-frequency domain, $\textbf{G}$ is the composition of all previous feed forward operations to get to that final fully connected layer, that is, $\textbf{G}(\mathcal{W}(S_n)) \in \mathbb{R}^{K \cdot T \cdot 2M_{\text{out}}}$, $\hat{\underline{\mathcal{W}}}(X_n)$ is the flattened output of the model which is an estimate of the true 12-lead ECG $X_n$ in the time-frequency domain, and $W_L \in \theta_d$ is our last layer weight matrix. $W_L$ is a square matrix with dimension $K \cdot T \cdot 2M_{out} \times K \cdot T \cdot 2M_{out}$ where we recall that $K$ is the number of frequencies and $T$ is time. From (\ref{eq: last layer}) we see that $\hat{\underline{\mathcal{W}}}(X_n)$, that is, the reconstructed ECG, is a linear combination of the columns of the $W_L$ matrix. In fact, the vector $\textbf{G}(\mathcal{W}(S_n))$ is giving the coordinate of the datum $S_n$ in the span of the columns of $W_L$. Therefore, in our formulation the ECG time-frequency space is spanned by the columns of $W_L$. Thus analyzing the columns of $W_L$ will provide insights into the $12$-lead ECG and what features make up this waveform. These learned features will be explored in detail below.

\subsection{Visualization}
\begin{figure*}
\centerline{\includegraphics[width=\textwidth, keepaspectratio]{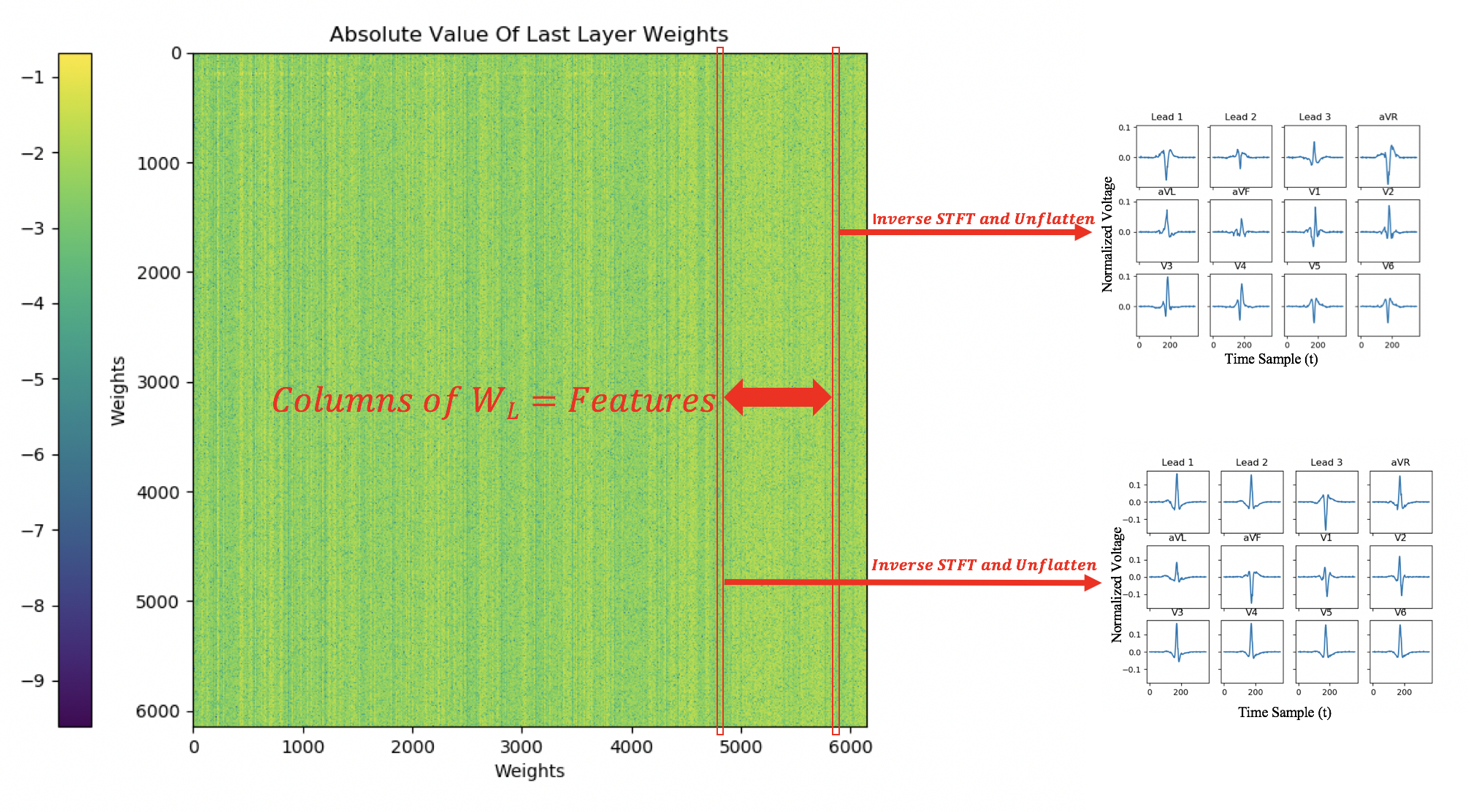}}
\caption{Last layer weight matrix and two reconstructed columns. The left panel is an image of our last layer weights in the model. The pixel values are log scaled for clearer visualization. Firstly, we notice that the weight matrix is highly sparse. This means that our learned features are localized in the time-frequency domain. This mimics how a $12$-lead ECG is also sparse when represented in the time-frequency domain. We can then select a column (red rectangle) and perform the inverse STFT to get back the feature in the time domain and then reshape the vector to match the dimension of $X_n$. We see that transforming these columns reveals ECG-like waveforms. Each column has slightly different morphologies and deflections as shown in the two traces on the right. For example, the feature on the top right of has a negative deflecting aVR lead while the bottom right feature has a positive deflecting aVR lead.}
\label{fig: Visualize}
\end{figure*}

Firstly we would like to visualize our $W_L$ matrix to get an idea for what is learned in the model. Furthermore, since each column of the matrix is the same dimension as our 12-lead ECG $\mathcal{W}(X_n)$, we can perform the inverse STFT to get back $X_n$ and visualize the features of $W_L$ in the time domain as in Fig. \ref{fig: forwardreconstruction}. In Fig. \ref{fig: Visualize}, we are displaying the last layer weights when the model is trained on half of the patients. Interestingly this matrix is highly sparse, indicating that our learned features are highly localized in the time-frequency domain. Note that our ECG and EGM waveforms were also highly sparse when converted into the time-frequency domain, see Fig. \ref{fig: complex}. Next, we performed the inverse STFT on the columns of the weight matrix and reshaped the output to get back a tensor of equal dimension to $X_n$. Interestingly these reshaped and transformed columns of $W_L$ resemble our $12$-lead ECG's. However, they exhibit slightly different morphologies depending on the column chosen for visualization. For example, the feature on the top right of Fig. \ref{fig: Visualize} has a negative deflecting aVR lead while the bottom right feature has a positive deflecting aVR. These observations lead to the hypothesis that the columns of our weight matrix $W_L$ form an overcomplete basis of our ECG space, with each column capturing slightly different morphologies exhibited in our training data.

\subsection{Ridge Regression for Reconstruction}
If the columns of our weight matrix do form a basis for our 12-lead ECG then we can verify that empirically by using multivariate linear regression to combine our features and reconstruct our patient's ECG data. To formalize this, we can say that any sample ECG can be expressed as

\begin{equation}
\hat{\underline{\mathcal{W}}}(X_n)=W_L \cdot \beta.
\label{eq: MultiLS}
\end{equation}
$W_L$ is again the last layer weight matrix and $\beta$ is our multivariate linear regression coefficients that tell us how to combine our features. Note that this also provides an alternative interpretation as to how our model operates. The NRCED transforms our EGMs into the same dimension as our 12-lead ECG and is solving for the proper coefficients to combine the features in $W_L$. So $\textbf{G}(\mathcal{W}(S_n))$ as defined in (\ref{eq: last layer}) is functionally equivalent to $\beta$;  $\textbf{G}(\mathcal{W}(S_n))$ is determined through composition of nonlinear operations performed on the EGM’s while $\beta$ is solved for using linear regression. We can then use (\ref{eq: rho}) to measure the correlation coefficient between our actual ECG heartbeats and our multivariate linear regression reconstructed heartbeats, similar to the procedure in (\ref{eq: loss}).

The solution to $\beta$ is a well known result in machine learning so the proof will not be shown \cite{hoerl1962applications}. However, the setup and closed form solution is provided below. Also we added an $L_2$ penalty to our coefficients, commonly known as ridge regression \cite{hoerl1962applications, hoerl1970ridge}. This is a necessary addition because our weight matrix is highly sparse and contains correlated columns, which makes the pseudo-inverse prone to diverging values. This mathematical formulation yields

\begin{equation}
\min_{\beta} ||\underline{\mathcal{W}}(X_n)-W_L \cdot \beta||_2^{2} + \lambda||\beta||_2^{2}
\end{equation}

\begin{equation}
\beta = (W_L^T \cdot W_L + \lambda \cdot I)^{-1}W_L^T \cdot \underline{\mathcal{W}}(X_n)
\end{equation}
Using this formulation allows a closed form solution for $\beta$, with the only parameter being $\lambda$. The best value for $\lambda$ was cross validated and found to be $10^{-7}$ for this problem. Equipped with $\beta$, we can now reconstruct 12-lead ECG's and determine if the features learned in our $W_L$ matrix are descriptive of the ECG waveform. 

\begin{figure}
\centerline{\includegraphics[width=\columnwidth,keepaspectratio, trim=4 4 4 4, clip]{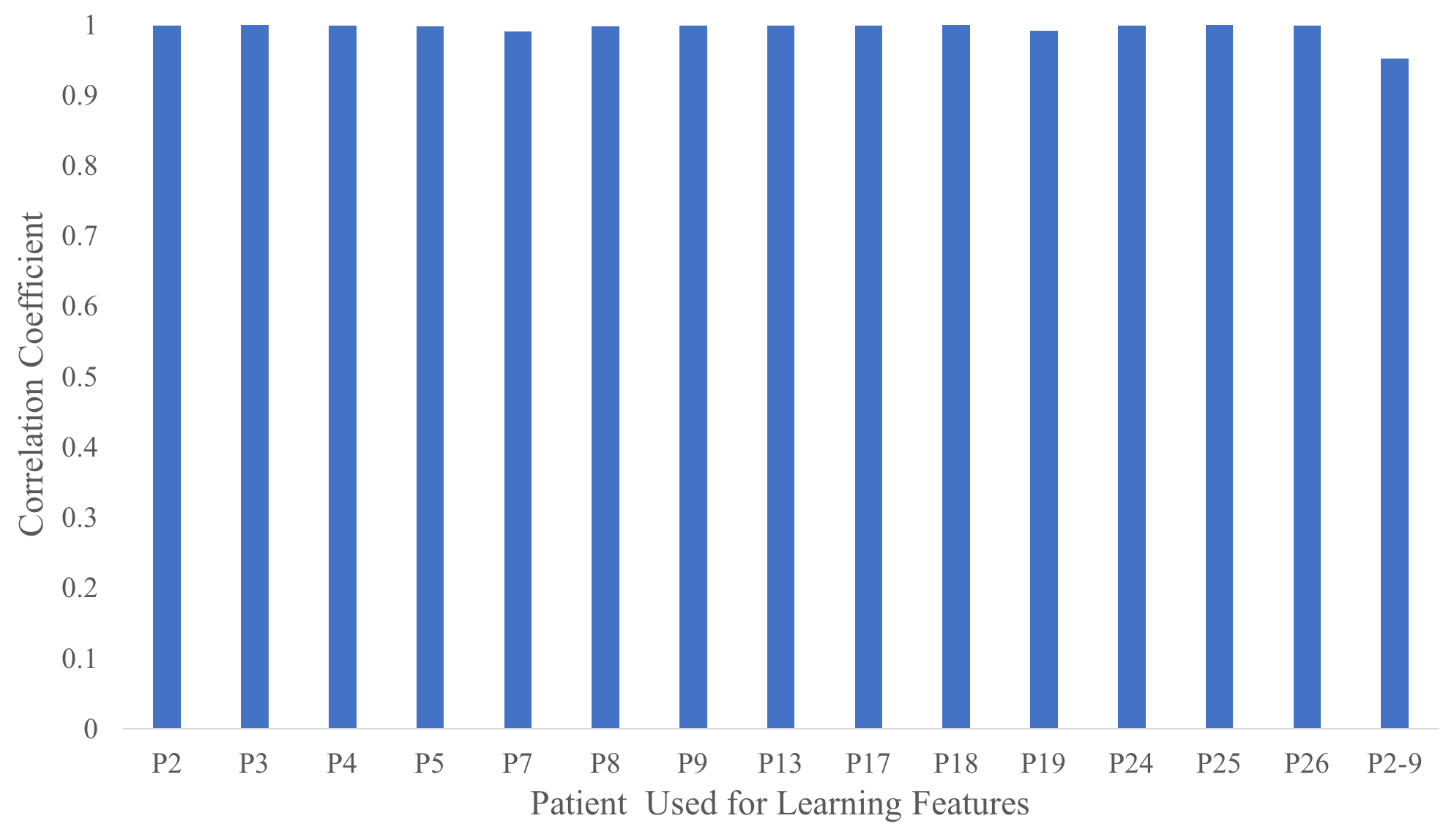}}
\caption{Correlation between reconstructed and original ECG using multivariate linear regression with different patient weights. We achieve nearly perfect reconstruction on all heartbeats when we use a weight matrix trained on only one patient. If we use a weight matrix trained on a group of patients we see a decrease in reconstruction capability. 
}
\label{fig: ridgecorr}
\end{figure}

The reconstruction was done by training the model on a specific patient's data or a set of patients, and then saving the final layer weights. For each $W_L$ matrix we reconstruct all $39,868$ heartbeats according to  (\ref{eq: MultiLS}) and calculate the correlation coefficient between our reconstruction and true 12-lead ECG as in (\ref{eq: rho}). The average was taken across all heartbeats and is reported in Fig. \ref{fig: ridgecorr}. We observe that a $W_L$ matrix trained on only one patient was able to achieve nearly perfect reconstruction of all heartbeats across patients. This is a surprising result since each patient had different ECG traces due to differences in heart physiology and pathology. Yet the features learned in $W_L$ were expressive enough to capture the idiosyncrasies of all heart beat morphologies regardless of which patient we trained on. Another observation is that if we used several patients to train $W_L$ we saw a decrease in reconstruction ability. This is most likely due to the fact that $W_L$ grows even sparser as we train on more patients. The features become less noisy, but they subsequently lose diversity in waveform morphologies.

\subsection{Identifying Atypical Heartbeats}
\begin{figure*}
\centerline{\includegraphics[width=\textwidth,keepaspectratio]{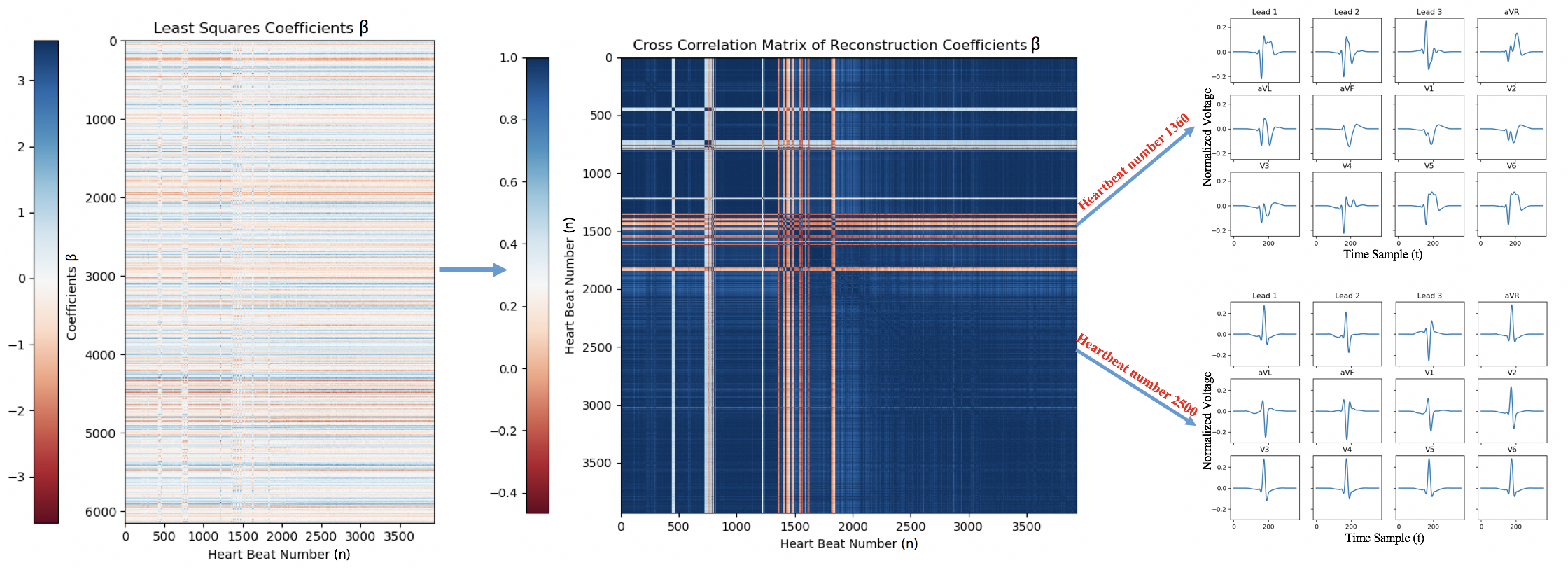}}
\caption{Visualization of $\beta$ when using a weight matrix trained on patient $5$ and its use for identifying atypical heartbeats on patient $7$. On the left we see a visualization of our beta coefficients from the multivariate linear regression. If we scan the image from left to right we see that the coefficients are generally the same except for distinct bands of heartbeats that require different coefficients in order to be reconstructed. To quantify this observation we calculated the cross correlation matrix, shown in the middle. Each element of our cross correlation matrix is equivalent to $\rho(\beta_{n_1}, \beta_{n_2})$, where $\beta_{n_1}$ is our linear regression coefficients for heart beat number $n_1$ and $\beta_{n_2}$ is our linear regression coefficients for heart beat number $n_2$. This is done for all combinations of $n_1$ and $n_2$. Since the Pearson correlation coefficient is a symmetric function, our cross correlation matrix is also symmetric. This matrix allows you to pick a heart beat and see how it compares to all other heartbeats in the recording. Blue indicates a strong positive correlation between two heartbeats while red indicates a strong negative correlation. We hypothesized that the rows that are majority blue are sinus rhythm heartbeats and the rows that are majority red are atypical or diseased heartbeats. We validated this by using the row number as our value for $n$ and plotting the associated $X_n$. We indeed found sinus rhythm heartbeats for high positive correlations indicated as blue rows (bottom right panel) and diseased heartbeats for negative correlations indicated as red rows (top right panel).  
}
\label{fig: atypical}
\end{figure*}

For each heartbeat we must solve for the particular $\beta$ that ensures proper reconstruction. This means that every heartbeat has its own set of coefficients. However, most of the heartbeats are sinus rhythm heartbeats so if you plot $\beta$ for every heartbeat you see the same pattern for a majority of the heartbeats. This can be seen on the left side of Fig. \ref{fig: atypical}. 
There are also bands where we see a difference in the values of $\beta$. This should indicate that non sinus heartbeats are occurring in the recording. We can take the cross correlation matrix to quantify how different these heartbeats are from the other heartbeats in the recording and to more clearly visualize when these heartbeats occur. Each element of our cross correlation matrix is equivalent to $\rho(\beta_{n_1}, \beta_{n_2})$ as in (\ref{eq: rho}), where $\beta_{n_1}$ is our linear regression coefficients for heart beat number $n_1$ and $\beta_{n_2}$ is our linear regression coefficients for heart beat number $n_2$. This is done for all combinations of $n_1$ and $n_2$. Since the Pearson correlation coefficient is a symmetric function, our cross correlation matrix is also symmetric. In this cross correlation matrix we see rows that vary in color from blue to red. Blue indicates a strong positive correlation between two heartbeats while red indicates a strong negative correlation. We hypothesized that the rows that are majority blue are sinus rhythm heartbeats and the rows that are majority red are atypical or diseased heartbeats. On the right side of Fig. \ref{fig: atypical} we can see a heart beat located on a red band in our correlation matrix and a heart beat located on a blue band. This is done by using the row number as our value for $n$ and plotting the associated $X_n$. Indeed, we see sinus rhythm heartbeats when the correlation is high and positive, as well as diseased heartbeats when the correlation is negative. Fig. \ref{fig: atypical} was done using $W_L$ trained on patient $5$, but this trend of finding clusters of diseased heartbeats holds for $W_L$ trained from any patient.

This result provides a way to identify atypical heartbeats from sinus rhythm heartbeats while a 12-lead ECG is being obtained. Not only do we identify that a beat is atypical, we can also assign a scalar value between $1$ and $-1$ as to how different that beat is from the average sinus beat (due to the definition of the Pearson correlation coefficient). Lastly, the heartbeats were ordered in time so the center panel of Fig. \ref{fig: atypical} also provides an idea of when these atypical episodes occurred and their duration. These extra pieces of information can help the cardiologist identify possible pathologies present in the patient.

\begin{figure}
\centerline{\includegraphics[width=\columnwidth,keepaspectratio]{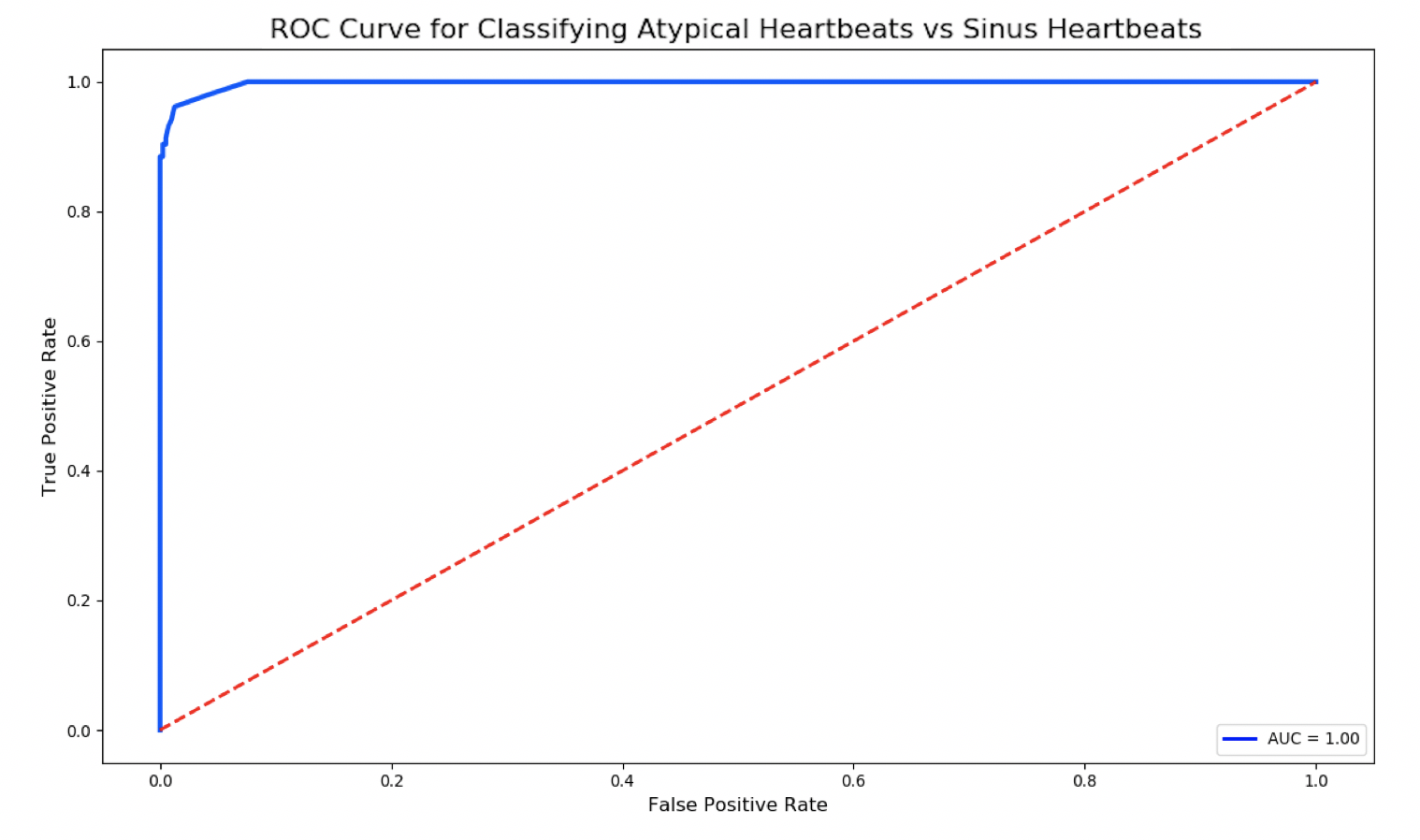}}
\caption{A receiver operating characteristic curve determined from the binary classification of sinus rhythm heartbeats vs atypical heartbeats for patient $7$. The associated area under the curve is approximately 1 indicating perfect classification between the two classes when using the entries of our cross correlation matrix as a discriminator.
}
\label{fig: ROC}
\end{figure}

To quantify the efficacy of this method we hand-labeled every heartbeat for patient $7$ with a $0$ for a sinus rhythm heartbeat and a $1$ for an atypical heartbeat. This was done with the aid of a cardiologist from the Texas Heart Institute. We then used the entries of our cross correlation matrix to predict the label of the heartbeat as $0$ or $1$ for varying threshold values. First, we rescaled the entries of our cross correlation matrix to be between $0$ and $1$ for simpler classification. Then, for a given heartbeat $n$ we assign that heartbeat a value $1$ if $\rho(\beta_{n}, \beta_{1}) < threshold$, where $\beta_{1}$ is the least square coefficients for a sinus rhythm heartbeat. With the true labels and predicted labels for varying thresholds we calculate a ROC curve \cite{lusted1971decision}. This is a common performance measurement for classification tasks that plots the true positive rate (also known as sensitivity) vs the false positive rate (also known as $1-$specificity) and is used extensively in medical diagnostic settings \cite{kumar2011receiver}. The ROC curve for this binary classification task is show in Fig. \ref{fig: ROC} and has an associated AUC value of approximately $0.98$. This indicates that our method of identifying atypical heartbeats vs sinus rhythm heartbeats has high accuracy, sensitivity, and specificity.

\section{Discussion}
\label{sec:Discussion}
This paper proposes and implements a methodology to synthesize a standard 12-lead ECG from a set of EGM leads, which is of practical importance in cardiology and electrophysiology. To achieve this, we built a nonlinear transformation between images of different channel number via our NRCED model. From an algorithm perspective the use of an Encoder-Decoder on spectrograms is not novel, but the use of an Encoder-Decoder to learn a nonlinear function between two different signals is novel. This has the potential to give Encoder-Decoders a new branch of applications for learning nonlinear functions between multivariate time series. 

The ability to build a nonlinear function between the ECG and EGM reveals some new insights about the signals. Since we can reconstruct ECG from EGM and vice versa then we can confidently say that the two signals share a low dimensional latent space. This gives confidence that in the case when we only have EGM’s and wish to view the concurrent 12-lead ECG we will be able to as long as the model is trained properly. The reverse should also be true, meaning if we have the 12-lead ECG and desire to know what the concurrent EGM’s look like we can reconstruct that as well. It was also surprising to see that we could reconstruct a full 12-lead ECG from a single EGM lead. This suggests that both the set of EGM’s and the 12-lead ECG contain redundant information that can be more efficiently expressed in a lower dimensional signal. This provides an explanation as to why the algorithm used in this paper worked so well for this problem since the model assumes a low dimensional representation shared between the input and output signal.

All literature in this field of ECG reconstruction has never attempted to perform the reverse reconstruction as we have shown. This capability arises naturally from the symmetric structure of the NRCED model. Using the algorithm in the reverse direction is straight forward and does not require an entirely new formulation to achieve. However, it does require us to train a slightly different model than the forward direction model. This method provides a non-invasive way of viewing a patients intracardiac electrograms, which was not possible prior to this work, as long as we have the 12-lead ECG. As mentioned above, this could be greatly useful in the context of cardiac ablation procedures to further isolate the problematic location of the heart before the procedure takes place.

The final layer of our model provided a way to interrogate the functionality of the algorithm and revealed a set of features that resembled a 12-lead ECG. These features were so descriptive of a 12-lead ECG that we could reconstruct all of our data with them via multivariate linear regression. The coefficients solved for in the multivariate linear regression provided a clever way to recognize atypical heart beats in a patient and ascribe a quantity as to how atypical the beat was from sinus rhythm. This could be used by a cardiologist to see when in time these episodes of atypical behavior occurred and for how long. It could also potentially be used to identify different sets of atypical behavior since the correlation matrix gives a continuous scalar from $-1$ to $1$. Perhaps diseased beats pertaining to different pathologies produce different correlation values, thus providing another interesting bit of information to the cardiologist. Another important use of this tool is as a way to construct better datasets for training the NRCED model. It is more useful to be able to reconstruct diseased heartbeats than sinus heartbeats. This is because we currently lack the ability to visualize a 12-lead ECG associated with a diseased set of EGM’s obtained on a pacemaker while a patient is away from the clinic. This indicates that we should train the NRCED on mostly diseased heartbeats so that we can confidently reconstruct the 12-lead ECG when the patient’s heart is in a diseased state. This requires us to build a training dataset of mostly diseased beats. This is difficult to do since the raw data is several hours long and contains thousands of heartbeats per patient. Searching for the diseased beats manually would take an incredibly long time and require an expert to identify. Using this new diagnostic tool would provide a quick and easy way to identify all atypical beats in a recording and then compile them into a training dataset. The AUC value of $0.98$ for the diagnostic tool gives high confidence that this process can be done. This means we can apply the NRCED to different heart pathologies and have a method of compiling appropriate datasets via the diagnostic tool.

\section{Conclusion and Future Work}
\label{sec:Conclusion}
This paper has shown a clear and effective application of a modified convolutional Encoder-Decoder named the NRCED to reconstruct a 12-lead ECG from a set of EGM leads. We first showed this result using five EGM leads as input, but showed identical success using only one EGM lead. We also proposed a generalized framework for using the algorithm such that you do not need to train the model on the patient of interest. Furthermore, we tested the model in the reverse direction to reconstruct EGMs from a 12-lead ECG and again saw high reconstruction accuracy. We also demonstrated that our model is interpretable and holds valuable information in the final layer weights. These weights were ultimately used to create a new diagnostic tool for identifying atypical and diseased heartbeats through time and to ascribe a scalar value as to how different these beats were from sinus beats. A highly desirable future direction would be to implement the NRCED algorithm on a custom chip by compressing the model and leveraging innovations in efficient deep learning. This provides a real route to application of the algorithm so that we may be able to constantly monitor a patient's cardiac health outside of the clinic and improve the quality of life for millions of people worldwide with pacemakers.

\section{Conflict of Interest Statement}
The authors of this paper have no conflicts of interest to report.

\section{Acknowledgement}
The results shown here are based upon data acquired with the approval of the Institutional Review Board for Baylor College of Medicine and Affiliated Hospitals with protocol number H-43925 and approval date August 22, 2018. This work has been supported by the National Heart Lung and Blood Institute under grant number R01HL144683.

\bibliography{main}
\bibliographystyle{ieeetr}

\end{document}